\newcommand{\beq}{\begin{eqnarray}}
 \newcommand{\eeq}{\end{eqnarray}}
\newcommand{\be}{\begin{equation}}
 \newcommand{\ee}{\end{equation}}

 \def\la{\mathrel{\mathpalette\fun <}}

\def\fun#1#2{\lower3.6pt\vbox{\baselineskip0pt\lineskip.9pt
\ialign{$\mathsurround=0pt#1\hfil ##\hfil$\crcr#2\crcr\sim\crcr}}}

\newcommand{{\SD}}{\rm SD}

\newcommand{\lan}{\langle}
\newcommand{\ran}{\rangle}

\documentclass{elsart}

\usepackage{amssymb}

\usepackage{graphicx}
\usepackage{amsmath}
\usepackage{amscd}
\usepackage{amsfonts}

\begin{document}

\begin{frontmatter}



\title{Nonperturbative equation of state of quark-gluon plasma. Applications}


\author{E.V.Komarov}
\ead{bartnovsky@itep.ru}
\author{Yu.A.Simonov}
\ead{simonov@itep.ru}

\address{State Research
Center Institute of Theoretical and Experimental Physics, Moscow,
117218 Russia}

\begin{abstract}
The vacuum-driven nonperturbative factors $L_i$ for quark and
gluon Green's functions are shown to define the nonperturbative
dynamics of QGP in the leading approximation. EoS  obtained
recently in the framework of this approach is compared in detail
with known lattice data for $\mu=0$ including $P/T^4$,
$\varepsilon/T^4$, $\frac{\varepsilon-3P}{T^4}$.  The basic role
in the dynamics at $T\la 3T_c$ is played by the factors $L_i$
which are approximately equal to the modulus of Polyakov line for
quark $L_{fund}$ and gluon $L_{adj}$. The properties of $L_i$ are
derived from field correlators and compared to lattice data, in
particular the Casimir scaling property $L_{adj} =(L_{fund})^{
\frac{C_2(adj)}{C_2(fund)}}$ follows in  the Gaussian
approximation valid for small vacuum correlation lengths.
Resulting curves for $P/T^4$, $\varepsilon/T^4$,
$\frac{\varepsilon-3P}{T^4}$ are in a reasonable agreement with
lattice data, the remaining difference points out to an effective
attraction among QGP constituents.
\end{abstract}

\begin{keyword}
nonperturbative thermodynamics \sep quark-gluon plasma
\PACS 12.38.Mh
\end{keyword}

\end{frontmatter}

\section{Introduction}
\label{}

Dynamics of Quark Gluon Plasma (QGP) is now of great interest,
since numerous results of heavy ion experiments call for strong
and possibly nonperturbative forces between quarks and gluons,
which cannot be explained  in the framework of perturbation
theory, see \cite{1} for reviews of recent results and their
interpretation.

Recently  one of the authors has proposed a new approach to the
study of the QGP dynamics \cite{2}, where the main emphasis was
done on the vacuum fields, and the resulting modification of quark
and gluon propagators was considered as the first and the basic
step in the nonperturbative (NP) treatment of  QGP, called Single
Line Approximation (SLA).

As a result one obtains  NP Equation of State (EoS) of QGP in  the
form of free quark and gluon terms multiplied by vacuum induced
factors. The latter are expressed via the only (nonconfining)
colorelectric correlator $D_1^E(x)$\cite{3}  and happened to be
approximately equal to the absolute values of Polyakov loops
$L_{fund}, L_{adj}$ for quarks and gluons respectively.

Thus all vacuum  NP dynamics in this approximation is encoded in
$L_{fund},$ and  $ L_{adj} =(L_{fund})^{9/4}$ by Casimir scaling
\cite{4}.

Moreover, the phase diagram was calculated in SLA \cite{5}
 assuming that the phase transition is again vacuum
dominated, {\it i.e.} a transition from confining vacuum with
vacuum energy density $\varepsilon_{conf} \cong
-\frac{\beta_0}{32} G_2(conf)$ to the nonconfining vacuum with
$\varepsilon_{dec}\cong -\frac{\beta_0}{32} G_2(dec)$.

The resulting phase curve $T_c(\mu)$ in \cite{5} depends on
$\Delta G_2=G_2 (conf) -G_2(dec)$ and  is in good agreement with
lattice data for standard values of $G_2 (conf)$
 \cite{6}  and $\Delta G_2\approx 0.35 ~G_2$ (conf).

 Thus the SLA is a reasonable starting point with no fitting or
 model parameters,  since $L_{fund}$ can be computed analytically
 \cite{3,7} or on the lattice \cite{8,9}, and $\Delta G_2$ is the  fundamental parameter of QCD \cite{5}.  This picture
 of the QCD phase transition was called in \cite{5} the Vacuum
 Dominance Model (VDM) originally proposed in \cite{10} in a
 simplified form (sometimes called the Evaporation Model).

In the model the basic element of the NP dynamics of QGP is the
quark and gluon Polyakov lines, which are connected to each other
by Casimir scaling. It is the purpose of this paper to study in
detail properties of Polyakov lines with the help of the Field
Correlator Method (FCM) \cite{11}  where those can be derived from
the nonvanishing colorelectric field correlator $D^E_1$. In
particular, $D_1^E(x)$ can be derived from the gluelump Green's
function, and the latter  was calculated analytically in
\cite{7,12} and on the lattice the gluelump spectrum was found in
\cite{13}. These properties can be compared to the lattice data
both at $T\leqslant T_c$ and $T>T_c$, and we predict behaviour of
$L_{fund}, L_{adj}$ at $T\leqslant T_c$  which   violates Casimir
scaling for $n_f=0$, since there $L_{fund}\equiv 0$ and
$L_{adj}=\exp \left(-\frac{m}{T}\right)$, with $m$ -- known
gluelump mass.  At $T \geqslant T_c$ Casimir scaling follows from
the dominance of quadratic (Gaussian) correlator, and we estimate
the admixture of higher correlators, violating the scaling.

At this point we notice, that contribution of bound states of
static quark or static adjoint charge with gluons in QGP to the
lattice defined $F_s(\infty, T)$ and consequently to $L_{fund},
L_{adj}$ would violate Casimir scaling, and the accurate
observation of Casimir scaling  in \cite{4} thus poses some limits
on those bound states.

As a result we fix the form of $L_{fund}, L_{adj}$ based on our
analytic and lattice calculations and enter with those to compute
EoS, {\it i.e.} $P(T), \varepsilon(T)$ and their derivatives.
Results of these computations are  compared with numerous lattice
data and shown to agree reasonably well within the accuracy of
lattice simulations.

The paper is organized as follows. In section 2 basic
thermodynamic equations for QGP are derived, and natural
appearance of Polyakov loops in EoS is derived. In doing so an
economic expression for the gluon pressure is first obtained,
while that of quarks is taken from \cite{2}.

In section 3 the expressions for $L_{fund}, L_{adj}$ are derived
in terms of field correlator $D^E_1$ and finally in terms of the
gluelump Green's function and properties of $L_{fund}, L_{adj}$
both below and above $T_c$ are discussed  in detail in comparison
with lattice data.

In section 4 EoS, $P(T)$ for $n_f=2, 2+1, 3,$ $\varepsilon (T)$
and nonideality $\frac{\varepsilon-3P}{T^4}$ are calculated using
the formulas of section 2 and compared to the lattice data.

Section 5 is devoted to the discussion of results and conclusions.

\section{Derivation of EoS for quark-gluon plasma}

Our derivation below is based on the formalism suggested in
\cite{2}, where the Background Perturbation Theory (BPTh) is
exploited, originally worked out in \cite{14} and developed in
connection to the FCM in \cite{15} for $T=0$ and in \cite{16} for
$T>0$. Correspondingly one splits the gluonic field $A_{\mu}$ into
background part $B_\mu$ and valence gluon part $a_\mu$, as \be
A_\mu=B_\mu+a_\mu \label{1}\ee and writes the partition function
$Z(B,T)$ as \be Z(B,T)=N\int D\phi \exp{(-\int_0^{\beta}dt \int
d^3x L_{tot}(x,t))}\label{2}\ee where $\phi$ denotes all set of
fields $a_\mu,\psi,\psi^+$ and ghost fields. In the lowest order
in $ga_\mu$ one obtains the result in the so-called Single Line
Approximation \be Z(B,T)=N_1
[det(G^{-1})]^{-1/2}det(-D_\lambda^2(B))[det(m_q^2-\hat{D}^2(B-\frac{i\mu_q}{g}\delta_{\mu
4}))]^{1/2}\label{3}\ee where $N_1$ is normalization constant,
$D_\lambda(B)=\partial_\lambda - igB_\lambda$, $G^{-1}=D_\lambda^2
\delta_{\mu \nu}+2igF_{\mu\nu}$. In what follows we put $\mu_q=0$,
and consider the case $\mu_q\neq0$ in a subsequent paper
\cite{17}.

The thermodynamic potential $F(T)$ is connected to $Z(B,T)$ in a
standard way \be F(T)=-T \ln\langle Z(B,T)\rangle_B\label{4}\ee
where the subscript $B$ in $\langle Z\rangle_B$ implies avaraging
over all background fields. As a result $F(T)$ in SLA is a sum of
gluon and quark degrees of freedom separately,
$F(T)_{SLA}=F_q(T)+F_{gl}(T)$. In what follows we omit the
subscript SLA, since all results (except for corrections to
Polyakov lines in the next section) will be valid in this
approximation. Using the Fock-Feynman-Schwinger (FFS) path
integral formalism (see \cite{18} for reviews) one has a
convenient representation
\begin{multline}
\frac{1}{T}F_{gl}(T)=Sp \left\{-\frac{1}{2}\int_0^\infty
\frac{ds}{s} \xi(s) e^{-sG^{-1}}+\int_0^\infty \frac{ds}{s}
\xi(s)e^{-sD^2(B)} \right\}=\\
-\int_0^\infty \frac{ds}{s}\xi(s)d^4x (Dz)_{xx}^w e^{-K} \left(
\frac{1}{2}tr\langle \hat\Phi_F(x,x)\rangle_B-\langle
tr\hat\Phi(x,x)\rangle_B\ \right) \label{5}
\end{multline}

Here \be \hat\Phi_F(x,y)=P_FP\exp{\left(ig\int_y^x B_\mu
dz^\mu\right)}\exp{\left(2ig\int_0^s F(z(\tau))d\tau\right)}
\label{6}\ee and $\hat\Phi(x,y)$ is the same as in (\ref{6})
without the last factor. $\xi(s)$ in the regularizing factor, for
details see \cite{2,16}. A similar representation for quarks and
antiquarks looks like \cite{16} \be
\frac{1}{T}F_q(T)=-\frac{1}{2}tr\int_0^\infty
\frac{ds}{s}d^4x\xi(s)(\overline{Dz})_{xx}^w e^{-K
-sm^2}W_\sigma(C_n)\label{7}\ee Note that in (\ref{5}),(\ref{7})
is present the "winding path measure", introduced in \cite{16},
e.g. for quarks
\begin{multline}
(\overline{Dz})_{xx}^w=\\
\lim_{N\rightarrow\infty}\prod_{n=1}^{\infty}\sum_{n=0}^{\infty}(-1)^n
\frac{d^4p}{(2\pi)^4}\exp{\left\{ip_\mu\left(\sum_{m=1}^\infty
\zeta_\mu(m)-(x-y)_\mu-n\beta\delta_{\mu4}\right)\right\}}
\label{8}\end{multline} And the same for gluons,
$(\overline{Dz})_{xy}^w$ but without the $(-1)^n$ factor. At this
point we are posing to contemplate the structure of our result
(\ref{5}),(\ref{7}) and recognize that it is a sum of individual
quark of individual quark or gluon lines (Green's functions in
background) over paths from $(\overrightarrow{x},0)$ to
$(\overrightarrow{x},n\beta)$.

It is clear that for $T<T_c$ this contribution should vanish
because of confinement, and one should look into the
representation containing gauge invariant Green's functions. These
come from white systems, e.g. for singlet $(gg)$ or $(q\bar q)$
and the corresponding partition function has the form \be
Z^{(n)}(\overrightarrow{x_1},\overrightarrow{x_2})=\int d\Gamma_1
d\Gamma_2 \langle tr W(C^{(1)}_n,C^{(2)}_n)\rangle \label{9}\ee
where $d\Gamma_i$ are phase space factors. Note the coinciding
indices in $W(C^{(1)}_n,C^{(2)}_n)$, which denotes the closed
Wilson loop (with possible insertions of $F_{\mu\nu}$ and
$\sigma_{\rho\lambda}F_{\rho\lambda}$ for quarks) starting at
points $(\overrightarrow{x_1},0)$, $(\overrightarrow{x_2},0)$
(connected by a parallel transporter) and ending at points
$(\overrightarrow{x_1},n\beta)$, $(\overrightarrow{x_2},n\beta)$
(again connected). Now, as was shown in \cite{2}, in the
deconfined phase the pair partition function factorizes in the
leading approximation of $(ga_\mu)^n$, while the color-electric
correlator $D_1^E$ yields nonzero contribution to each quark or
gluon in the form of Polyakov lines. The derivation is shortly as
follows (see \cite{2} for details).

The Wilson loop in (\ref{9}) can be calculated in terms of field
correlators using cluster expansion theorem \cite{19}

\begin{multline}
\frac{1}{N_c}tr \langle W(C^{(1)}_n,C^{(2)}_n)\rangle=\frac{1}{N_c}tr \langle P\exp{(ig\int_{C_n}B_\mu dz^\mu)}\rangle_B=\\
\exp{\left(-\sum_{k=2}^{\infty}\frac{(ig)^k}{k!}\int\int_{S_n}ds_{\mu_1\nu_1}(u_1)\ldots
ds_{\mu_k\nu_k}(u_k)D_{\mu_1\nu_1\ldots
\mu_k\nu_k}(u_1,\ldots,u_k)\right)} \label{10}
\end{multline}
where $C_n$ is the total closed loop, containing
$C^{(1)}_n$,$C^{(2)}_n$ and parallel transporters from
$\overrightarrow{x_1}$ to $\overrightarrow{x_2}$ and back, $S_n$
is a surface inside $C_n$, while the field correlators are defined
as follows, e.g. for $k=2$ (Gaussian approximation) one has \be
D_{\mu_1\nu_1\mu_2\nu_2}(u_1,u_2)=\frac{g^2} {N_c}tr\langle
F_{\mu_1\nu_1}\Phi(u_1,u_2)F_{\mu_2\nu_2}\Phi(u_2,u_1)\rangle
\label{11}\ee In what follows we concentrate on color-electric
correlators, which can be written in terms of two scalar functions
$D^E(w),D^E_1(w)$ (for contribution of other color-magnetic
correlators see \cite{2,20}. Note that latter do not produce
factorized contribution, but can support weakly bound states with
angular momentum $L>0$)
\begin{multline}
D_{i4,k4}(x,y)=\frac{g^2}{N_c}\langle tr E_i(x)\Phi(x,y)E_k(y)\Phi(y,x)\rangle=\\
\delta_{ik}(D^E+D_1^E+u_4^2\frac{\partial D_1^E}{\partial
u_4^2})+u_i u_k \frac{\partial D_1^E}{\partial u^2} \label{12}
\end{multline}
where $D^E\equiv D^E(u)$, $D_1^E\equiv D_1^E(u)$, $u=x-y$.

We now take into account according to \cite{7,12} that correlation
lengths $\lambda^E$ and $\lambda^E_1$, defined from asymptotics
$D^E(u)\sim \exp{(-|u|/\lambda^E)}$, $D^E_1\sim
\exp{(-|u|/\lambda^E_1)}$, are small, $\lambda^E$,
$\lambda^E_1<0.2$ fm. Indeed from the gluelump correlators
\cite{12,13} it follows that $\lambda^E\approx 0.08$ fm,
$\lambda^E_1\approx 0.16$ fm. Then for temperatures
$T<1/\lambda^E$, $1/\lambda^E_1$ the $n$ dependence appears
explicitly \cite{2} and one can write \be \frac{1}{N_c}tr\langle
W(C^{(1)}_n,
C^{(2)}_n)\rangle=\exp{(-w^{(2)}_n-w^{(4)}_n-\ldots)}\label{13}\ee
where the Gaussian contribution is expressed via $D^E$, $D^E_1$
\be w^{(2)}_n=n\beta \left(V_D(r,T)+V_1(r,T)\right)\label{14}\ee
and we have defined \cite{2,3} \be V_D(r,T)=2 \int_0^\beta d\nu
(1-\nu T)\int_0^r (r-\xi)d\xi D^E(\sqrt{\xi^2+\nu^2})\label{15}\ee
\be V_1(r,T)=2 \int_0^\beta d\nu (1-\nu T)\int_0^r \xi d\xi
D^E_1(\sqrt{\xi^2+\nu^2})\label{16}\ee Here
$r=|\overrightarrow{x_1}-\overrightarrow{x_2}|$.

Now it is clear, that in the confined regime, when $D^E$ is
nonzero, $V_D(r,T)$ grows linearly with $r$, and factorization of
$Z^{(n)}(C^{(1)}_n, C^{(2)}_n)$ is impossible - $gg$ and $q\bar q$
propagate as hadrons. For $T>T_c$, however, $D^E\equiv 0$ and \be
V_1(r,T)=V_1(\infty,T)+v(r,T)\label{17}\ee where
$v(r,T)|_{r\rightarrow \infty}=0$ and contains both perturbative
and NP contributions. In \cite{9} it was shown that $v(r,T)$ is
able to support weakly bound states of heavy quark and antiquark
$(Q\bar Q)$, as well as (gg) and $Qg$ systems. This is in
agreement with lattice data \cite{21}. In the SLA approximation we
neglect in the first step the effect of $v(r,T)$ and keep only
$V_1(\infty,T)$. As will be shown below this latter contribution
explains EoS of QGP with good accuracy. Then the gauge invariant
quark-antiquark Green's function factorizes into a product of
one-body terms, each obtaining a factor \be
L_{fund}^{(n)}\cong\exp{(-n
\frac{V_1(\infty,T)}{2T})}=\left(L_{fund}^{(1)}\right)^n\label{18}\ee
For the gluon $gg$ system one obtains in addition in the exponent
the Casimir factor $\frac{C_2(adj)}{C_2(fund)}=\frac{9}{4}$, which
follows from (\ref{10})-(\ref{12}), when all fields are in the
adjoint representation, \be L_{adj}^{(n)}=\exp{(-n
\frac{9V_1(\infty,T)}{8T})}\label{19}\ee
 In the next section we study these factors in more detail and establish their relation to the Polyakov
 loop factors measured on the lattice. We end this section with the discussion of higher correlators in (\ref{10}).

Keeping for smooth surfaces only even power correlators (see
discussion in \cite{22}) one can estimate the contribution of the
$k=4$ correlator compared to the Gaussian one in the exponent of
(\ref{10}) as giving additional factor \be \eta =
\frac{w^{(4)}_n}{w^{(2)}_n}=\overline{(gF)^2}(\lambda^E)^4\approx
\sigma_E(\lambda_E)^2\sim \frac{0.2 \mbox{ GeV}^2}{(2 \mbox{
GeV})^2}<0.1\label{20}\ee where in confinement phase we have
estimated $\overline{(gF)^2}$ from the string tension
$\sigma_E=\frac{1}{2}\int d^2x D^E(x)\approx \overline{(gF)^2}\
\overline{(\lambda^E)^2}\approx 0.2\mbox{ GeV}^2$. Note that
estimate of $\overline{(gF)^2}$ from the gluonic condensate yields
$\eta$ order of magnitude smaller. The estimate (\ref{20}) gives a
reasonable explanation of the good accuracy of Casimir scaling in
the confined phase (see \cite{23} for discussion). In the case of
deconfinement, when $\sigma_E=0$ and gluonic condensate is roughly
twice as small (up to $T\approx 1.5 T_c$ \cite{8}) the "Casimir
expansion parameter" $\eta$ should be even smaller, since
$\lambda_1^E$ does not change significantly \cite{8}, while
$\lambda^H_1$, $\lambda^H$ stay constant in this
region.\footnote{Note that estimate (\ref{20}) refers to the plane
or at least smooth surface, while for a crumpled surface higher
correlators play important role to make area law with the minimal
surface.}

We obtain doing path integrals as explained in \cite{2}, \be
P_{gl}=\frac{N_c^2-1}{16\pi^2}\int_0^\infty
\frac{ds}{s^3}\sum_{n\neq
0}e^{-\frac{n^2\beta^2}{4s}}L_{adj}^{(n)}\label{21}\ee \be P_q=n_f
\frac{N_c}{4\pi^2}\int_0^\infty \frac{ds}{s^3}e^{-m_q^2 s}
\sum_{n=1}^{\infty}(-1)^{n+1}e^{-\frac{n^2\beta^2}{4s}}L_{fund}^{(n)}\label{22}\ee
Finally performing integration over $ds$ one has \be
P_{gl}=\frac{2(N_c^2-1)}{\pi^2}T^4\sum_{n=1}^{\infty}\frac{1}{n^4}L_{adj}^{n}\label{23}\ee
\be P_q=\frac{4N_c n_f}{\pi^2}T^4\sum_{n=1}^{\infty}
\frac{(-1)^{n+1}}{n^4}L_{fund}^{n} \varphi_q^{(n)} \label{24}\ee
where we have defined $L_{adj}\equiv L_{adj}^{(1)}$,
$L_{fund}\equiv L_{fund}^{(1)}$, see (\ref{18}),(\ref{19}), and
\be \varphi_q^{(n)}=\frac{n^4}{16T^4}\int_0^\infty
\frac{ds}{s^3}e^{-m_q^2s - \frac{n^2\beta^2}{4s}}=\frac{n^2
m_q^2}{2T^2}K_2(\frac{n m_q}{T})\label{25}\ee These equations and
another, integral form instead of the infinite sum, will be used
in section $4$ to compare with lattice data.

\section{Polyakov lines and field correlators}

  Below only quadratic (Gaussian) field correlators  are
  considered, basing on the Casimir scaling property which these
  correlators ensure, and being in agreement with lattice data
  both for $T=0$ \cite{23} and for $T>T_c$ \cite{4}. At $T>0$ four
  Gaussian correlators   are $D^E(x), D_1^E(x), D^H(x), D_1^H(x)$,
  with $\sigma^{E,H}=\frac12 \int D^{E,H} (x) d^2 x$. At $T>T_c$
  the correlator $D^E$ and $\sigma^E$ vanish, as was suggested in
  \cite{10} and proved on the lattice \cite{8}, and three other
  correlators are nonzero,   moreover the spatial string tension
  $\sigma_s\equiv\sigma^H$ grows with temperature in the dimensionally reduced limit \cite{24}.
  This fact explains also the growth with temperature of the Debye mass, $m_D\cong 2
  \sqrt{\sigma_s}$ \cite{25}, which is known from lattice data
  \cite{24}. Apart  from this  quantity, we shall not use  below
  the  colormagnetic correlators, since they do not produce static
  potentials for interparticle angular momentum
  $L=0$.

  Therefore we shall be interested only in color-electric (CE)
  correlators  $D^E(x)$ (inside the confining phase bounded by the
  curve $T_c(\mu))$, and $D_1^E(x)$ in the whole $\mu, T$ plane.

  It is important at this point to stress that in our approach only gauge
  invariant states $|n\ran$ are to be considered in the partition
  function at $T>0$, \be
  Z=\sum_n \lan n|e^{-H/T}|n\ran\label{26}\ee
  as  well as in all QCD states at $T=0$. This is evident in the
  confining phase, since a colored  part
  of the gauge invariant system is connected by the  string to
  other parts.

  With the lack of string   in the deconfined phase the necessity
  of using the gauge invariant amplitudes is less evident, except
  for worldlines in the  spatial directions, where colormagnetic
  confinement with nonzero $\sigma_s$ is operating.

  Nevertheless our use of gauge invariant amplitude, which
  factorizes at large interparticle distances in the deconfined
  phase, leads to the explicit prediction of EoS with modulus
  of phase factors, which approximately equal to modulus of
  Polyakov lines.

  Below we shall use, as in \cite{2,5}, the gauge invariant states, $|n\ran$
  at all $\mu, T$ and we shall  express the interparticle
  dynamics in  terms of gauge invariant quantities, like pair or
  triple static potentials. The large distance limit of these
  potentials yields one-particle characteristics -- the
  self-energy parts of quarks, antiquarks, gluons etc.  One can
  use those to study thermodynamics of QGP in the one-particle,
  or Single Line Approximation (SLA) \cite{2}. It is rewarding,
  that the field correlator method is a natural instrument in
  describing this deconfined dynamics, since in absence of $D^E$
  the correlator $D^E_1$ has the form of the full derivative and
  produces gauge invariant one-particle pieces -- self-energy
  parts -- automatically (in addition to interparticle interaction
  decreasing at large distances).

  The gauge invariant states $|n\ran, \lan n|$ formed with the help of
  parallel transporters (Schwinger lines)
  $\Phi(x,y)\equiv P\exp (ig \int^x_y A_\mu dz_\mu)$, create, as
  shown in \cite{2}, Wilson loops $W(C)$ for $q\bar q, qqq$, or
  else ($qq\bar q\bar q$) systems. From the latter, as shown in
  \cite{2,3}, one obtains static potentials. When treating colored
  systems like $(qq)$, the latter is taken as a part of  gauge
  invariant system ($qq\bar q\bar q$), and the pairs $(qq)$ and $\bar q\bar
  q$) are separated at large distance where potential $V(qq,\bar q\bar
  q$) is neglected.

  We start with the color singlet $q\bar q$ system and write
  contributions of $D^E, D^E_1$ at nonzero $T=1/\beta$ to the
  static potentials \cite{3}
  \be V_1(r,T)= \int^\beta_0
d\nu (1-\nu T) \int^r_0 \xi d \xi
D^E_1(\sqrt{\xi^2+\nu^2})\label{27}\ee \be V_D(r,T)= 2\int^\beta_0
d\nu (1-\nu T) \int^r_0 (r-\xi) d \xi
D^E(\sqrt{\xi^2+\nu^2})\label{28}\ee

It is important that $V_1, V_D$ give the contribution to the
modulus of Polyakov loops, namely \cite{3} \be L_{fund}^{(V)}
=\exp (-\frac{V_1
(T)+2V_D}{2T}),~~L_{adj}^{(V)}=(L_{fund}^{(V)})^{9/4}\label{29}\ee
where $V_1(T) \equiv V_1(\infty, T), ~~ V_D\equiv V_D(r^*, T)$ and
$r^*$ is an average distance between the heavy quark line and
light antiquark (for $n_f>0$) , or ``heavy gluon line'' and a
gluon for $L_{adj}$. The Casimir scaling relation (\ref{29})
predicted in \cite{3}  is in good agreement with lattice data
\cite{4}, as well as vanishing of $L_{fund}$ for $T\leqslant T_c,
n_f=0$ and the strong drop of $L_{adj}$ for $T\leqslant T_c$.
Indeed, for $T \leqslant T_c$ and $n_f=0$ one has $r^*\rightarrow
\infty$ and $V_D \rightarrow \infty$, explaining the vanishing of
$L_{fund}^{(V)}$. For $L_{adj}^{(V)}$ in this region one can take
into account the kinetic energy of the gluon in the system adjoint
source plus gluon in a gluelump. This yields an estimate
$L_{adj}(T\leqslant T_c)=\exp{(-\frac{m_{glp}}{T})}$, where
$m_{glp}$ was computed in \cite{12,13} to be $\approx 1$ GeV.

In \cite{3} it was mentioned, that Polyakov lines measured
repeatedly on the lattice, are expressed through the (singlet)
free energy of $Q\bar Q$ system at large distances $F^1_{Q\bar Q}
(\infty, T)$ in the same way as in (\ref{29}), i.e \be
L_{fund}^{(F)} =\exp \left( -\frac{F^1_{Q\bar Q} (\infty, T)}{2T}
\right)\label{30}\ee and actually the difference between
$L_j^{(F)}$ and $L_j^{(V)}$ was not taken into account in
\cite{3}. This
  difference can be easily seen in the standard representation of
  $F^1_{Q\bar Q}(r, T)$
  \be
  \exp \left(-\frac{F^1_{Q\bar Q} (r,T)}{T}\right)= \sum_{n(Q\bar
  Q)} c_n\exp \left(-\frac{V_n^{Q\bar Q}
  (r,T)}{T}\right)\label{31}\ee
  where $n(Q\bar Q)$ denote all excited and bound states where
  $Q\bar Q$ participate, and $V_n^{Q\bar Q}(r,T)$ is the energy term
  of such state $n$ when distance between static charges $Q$ and
  $\bar Q$ is equal to $r$. It is clear that $L_j^{(V)}$ coincides
  with $L_J^{(F)} $ when all states $n$ except for the ground
  state $n=0$ are neglected. In this case $V_0^{Q\bar Q}(r,T)$
  coincides with $V_1(r, T)$, and hence with $F^1_{Q\bar Q}
  (r, T)$.  Note at this point, that $V_1(r, T)$ in (\ref{27})
  does not depend on $T$ in the limit when the vacuum correlation
  length $\lambda (D_1^E(x) \sim e^{-x/\lambda})$ tends to zero,
  $T\ll\frac{1}{\lambda} \approx 1$ GeV.

  In the general case all states $n(Q\bar Q)$ contribute and
  therefore $(c_n>0)$ one has inequality
  \be
  V_1(r, T) \geqslant  F^1_{Q\bar Q} (r, T)\label{32}\ee

To define $V_1$ and $L_{fund}$ properly, one should separate
perturbative and NP parts and renormalize $V_1$ to get  rid of
perimeter divergences.

The separation in  $D^E_1(x)$ can be seen at small $x$ \cite{7}

\begin{multline}
D_1^E (x) =\\
 \frac{4C_2(f)\alpha_s}{\pi} \left\{
\frac{1+O(\alpha_s \ln^k x) }{x^4} + \frac{\pi^2G_2}{24 N_c} +
\ldots\right\} = D_1^{E ~pert} (x)+ D_1^{(np)} (x) \label{33}
\end{multline}
and at
large $x$, $D^{(np)}_1 (x) $ is \cite{7}. \be D_1^{(np)} (x) \cong
A_1 \frac{ e^{-M_0|x|}}{\sqrt{x^2}}, A_1=C(f) \alpha_s 2 M_0
\sigma_{adj}\label{34}\ee where $M_0$ is the lowest gluelump mass
\cite{12,13}, $M_0 \approx 1$ GeV.

The corresponding separation of  $V_1(r,T)$ is done in  \cite{3,9}
as follows \be V_1(r,T) = V_1^{pert} (r,T) + V_1^{(np)} (r,T) +
V_1^{(div)}(a) \label{35}\ee where \be V_1^{(pert)} (r,T)
=-\frac{C(f) \alpha_s}{r} e^{-m_Dr}(1+ O(rT))\label{36}\ee
$V^{(np)}$ is as in (\ref{27}) with $D^E_1\to D^{np}_1(x)$, \be
V_1^{(div)} (a) \cong \frac{2C(f) \alpha_s}{\pi} \left(
\frac{1}{a} +O(T\ln{a})\right) \label{37}\ee

Here $m_D=m_D (T) \approx  2\sqrt{\sigma_s}$ is the $np$ Debye
mass \cite{25},   and  $a$ is the lattice  cut-off.

\begin{figure}[t]
\includegraphics[width=12cm]{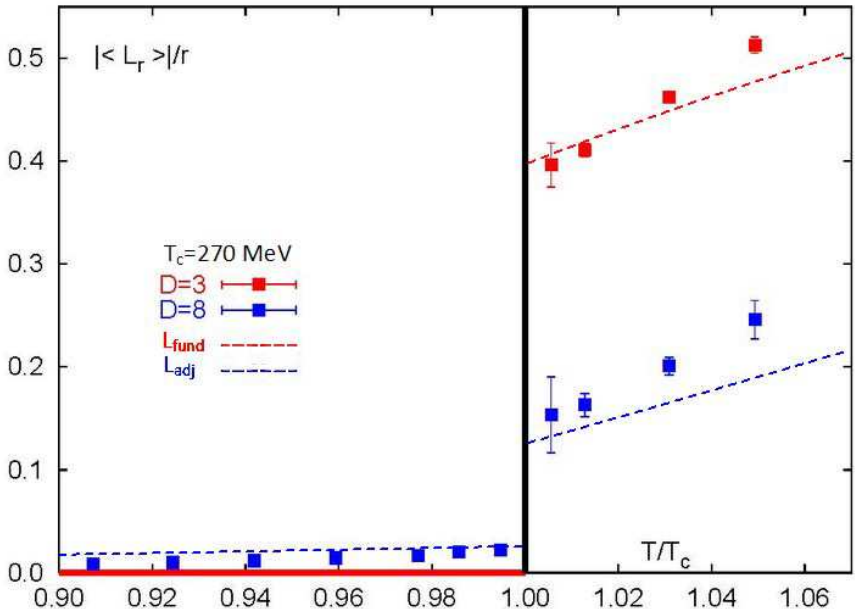}
\caption{Shown on the figure are curves of $L_{adj}$ (blue dashed)
and $L_{fund}$ (red dashed) compared to the ones taken from
\cite{4}. In the $T<T_c$ region the
$M(\bar{\alpha}_s=0.195)=0.982$ GeV gluelump mass was used. In the
deconfinement region the fit (\ref{39}) was used with $T_c=270$
MeV for $L_{fund}$ and the Casimir scaled value for $L_{adj}$.}
\end{figure}

The renormalization procedure suggested in \cite{3} amounts to
discarding $V_1^{(div)} (a)$, and this is in agreement with the
lattice renormalization used in \cite{26}, where $F^1_{Q\bar Q}(r,
T)$ was  adjusted to the form  $V_1^{pert}(r, T)$ at small $r$ and
$T$. Note, that $V_1^{np} (r,T)\sim O(r^2)$ in this region and the
procedure indeed allows to eliminate the constant term
$V_1^{div}(a)$.


We start with the one-particle limit of $V_1(r,T)$, and the
corresponding contribution to $L_{fund}^{(V)}$.

According to the discussion above, one defines the renormalized
Polyakov loop as in (\ref{29}),(\ref{35}) with $V_1(T)\equiv
V_1^{np} (\infty, T)$ and we shall neglect the difference between
$L_j^V$ and $L^{(F)}_j$ (important at large $T$, where
$L_{fund}^F>1$, while always $L_{fund}^V<1$).

From (\ref{34}) one has (at $T\leqslant T_c$) \be
V_1^{(np)}(\infty, T) =\frac{A_1}{M_0^2}\left[ 1- \frac{T}{M_0}
\left( 1-e^{-\frac{M_0}{T}}\right)\right]\label{38}\ee so that
$V_1^{(np)} (\infty, T_c)\approx \frac{6 \alpha_s (M_0)
\sigma_f}{M_0} \approx 0.5$ GeV for $M_0 \approx 1$ GeV
\cite{12,13}.

The same type of estimate one obtains from lattice data \cite{27}
where at $T\gtrsim T_c$ one can parametrize the data as follows
\be F^1_{Q\bar Q}(\infty, T) \approx \frac{0.175}{1.35 \left(
\frac{T}{T_c}\right) -1}, F^1_{Q\bar Q}(\infty, T_c) \approx 0.5
~{\rm GeV}\label{39}\ee

Thus one can say that quarks (and antiquarks) have selfenergy
parts $\kappa_q (T) =\kappa_{\bar q} (T) =\frac12 V_1 (T) \approx
\frac12 F^1_{Q\bar Q} (\infty, T) \approx 0.25$ GeV at $T\approx
T_c$.

To illustrate our discussion of $V_D$, $V_1$ and $L_{fund}$,
$L_{adj}$ we show in Fig.1 our curves for $L_{fund}$, $L_{adj}$
computed from (\ref{29}) with $V_1(\infty,T)=F_{Q\bar
Q}^1(\infty,T)$ taken from (\ref{16}) for $T > T_c$ while
$L_{adj}=\exp{(-\frac{M_0}{T})}$ for $T \leqslant T_c$. Our dashed
curves are plotted in Fig.1 in comparison to lattice data from
\cite{4}.

For gluons one has instead $\kappa_g (T) =\frac94 \kappa_q \approx
0.56$ GeV. Let us turn  now to the $r$-dependence of interaction.
The perturbative part has a standard screened Coulomb behaviour
(\ref{36}), while the NP part vanishes at small $r$; \be V_1^{np}
(r, T) \sim \mbox{const}\cdot r^2,~~ r\to 0\label{40}\ee

From (\ref{27}),(\ref{34}) one has as in \cite{3}
\begin{multline}
 V_1^{(np)}(r,T) = \\
 V_1^{(np)} (\infty, T) -\frac{A_1}{M^2_0} K_1 (M_0 r) M_0
r +O(\frac{T}{M_0})\equiv V_1^{(np)} (\infty, T) +v(r,T)
\label{41}
\end{multline}
Hence the NP interaction in the white system $Q\bar Q$ changes
from $V_1^{np} (\infty, T) \approx 0.5 $ GeV at large $r$ to zero
at small $r$. The same (multiplied by $\frac94$) is true for the
white $gg$ system.


We end up this section by discussion of the role of excited states
in definition of $F^1_{Q\bar Q}$ and possible violation of Casimir
scaling for $L_{fund}, L_{adj}$. It is clear that in $F^1_{Q\bar
Q}$ for $n_f=0$ the only possible excited states consist of gluons
$(Qg)(\bar Qg)$; $(Qgg)(\bar Qgg)$ etc. As it was shown in
\cite{9}, the weakly bound states $(Qg)$ indeed are supported by
$V_1(r,T)$, and neglecting the small binding energy the total
energy of these states is roughly the sum of selfenergy parts
$\kappa_Q$ and $\kappa_g$ \be
E_{Qg}\approx\frac{1}{2}V_1(\infty,T)+\frac{9}{8}V_1(\infty,T)\approx
0.8 \mbox{ GeV}(T\approx T_c)\label{42}\ee This should be compared
to the possible bound state of an adjoint static source $G$ plus
gluon, which in the weakly binding limit can be written as \be
E_{Gg}\approx 2\cdot \frac{9}{8}V_1(\infty,T)\approx1.1 \mbox{
GeV} (T\approx T_c) \label{43}\ee In addition multiplicities of
states (\ref{42}) and (\ref{43}) are different, which leads to
different predictions for corrections to $F^1_{Q\bar Q}$ and
$F^1_{GG}$, not connected by Casimir scaling, in contrast to the
main (ground state) term $V^1_{Q\bar Q}=V_1(\infty,T)$ and
$V^1_{GG}=\frac{9}{4}V_1(\infty,T)$. Therefore one expects
violation of Casimir scaling by gluon-induced bound states in
$L_{fund}$ and $L_{adj}$, and high accuracy of lattice data
\cite{4} indicates then a small role of such bound states.

\section{A comparison to the lattice data}

\begin{figure}[!h]
\includegraphics[width=6.5cm]{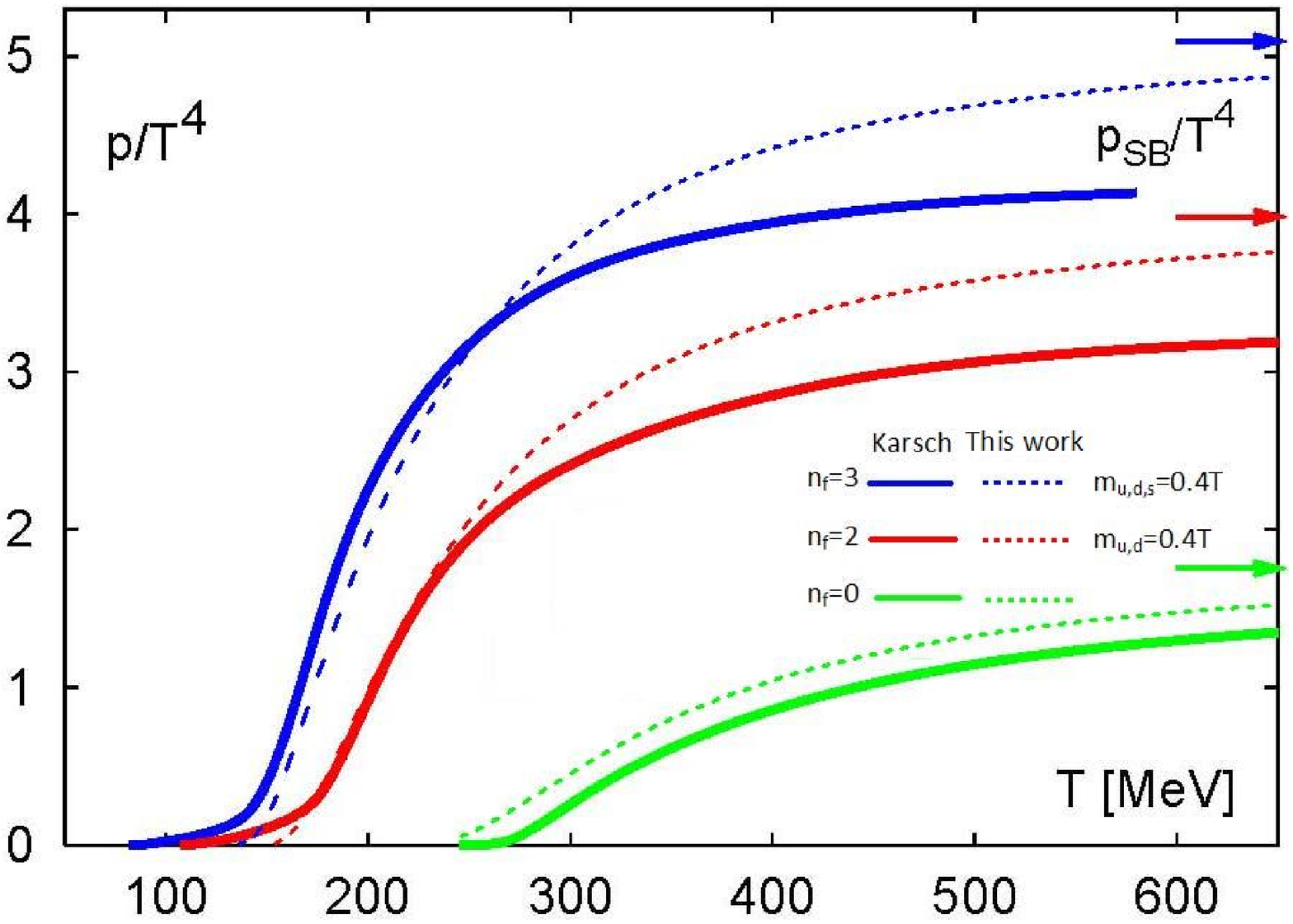}
\includegraphics[width=6.5cm]{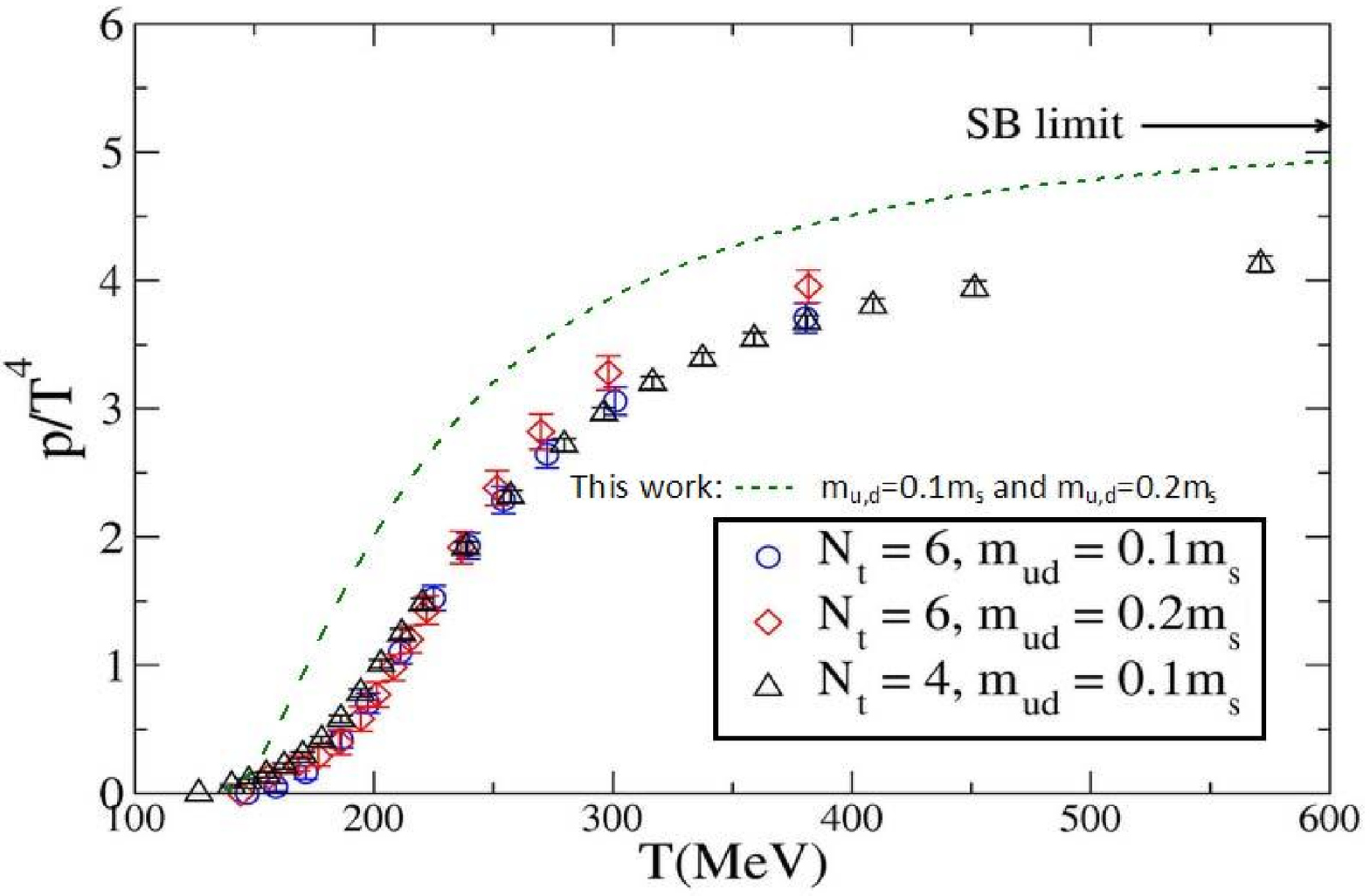}
\caption{Pressure $\frac{P}{T^4}$ as function of temperature $T$.
Shown on the left figure is a comparison of the analytical
calculus (\ref{48}),(\ref{49}) (dashed lines) with the lattice
results (bold lines) \cite{0608003} for the case $n_f=0,2,3$.
Shown on the right figure is the case of $n_f=2+1$. Green dashed
line is the analytical calculation (\ref{48}),(\ref{49}) compared
to the lattice one from \cite{0610017}.}
\end{figure}

\begin{figure}[!h]
\includegraphics[width=7cm]{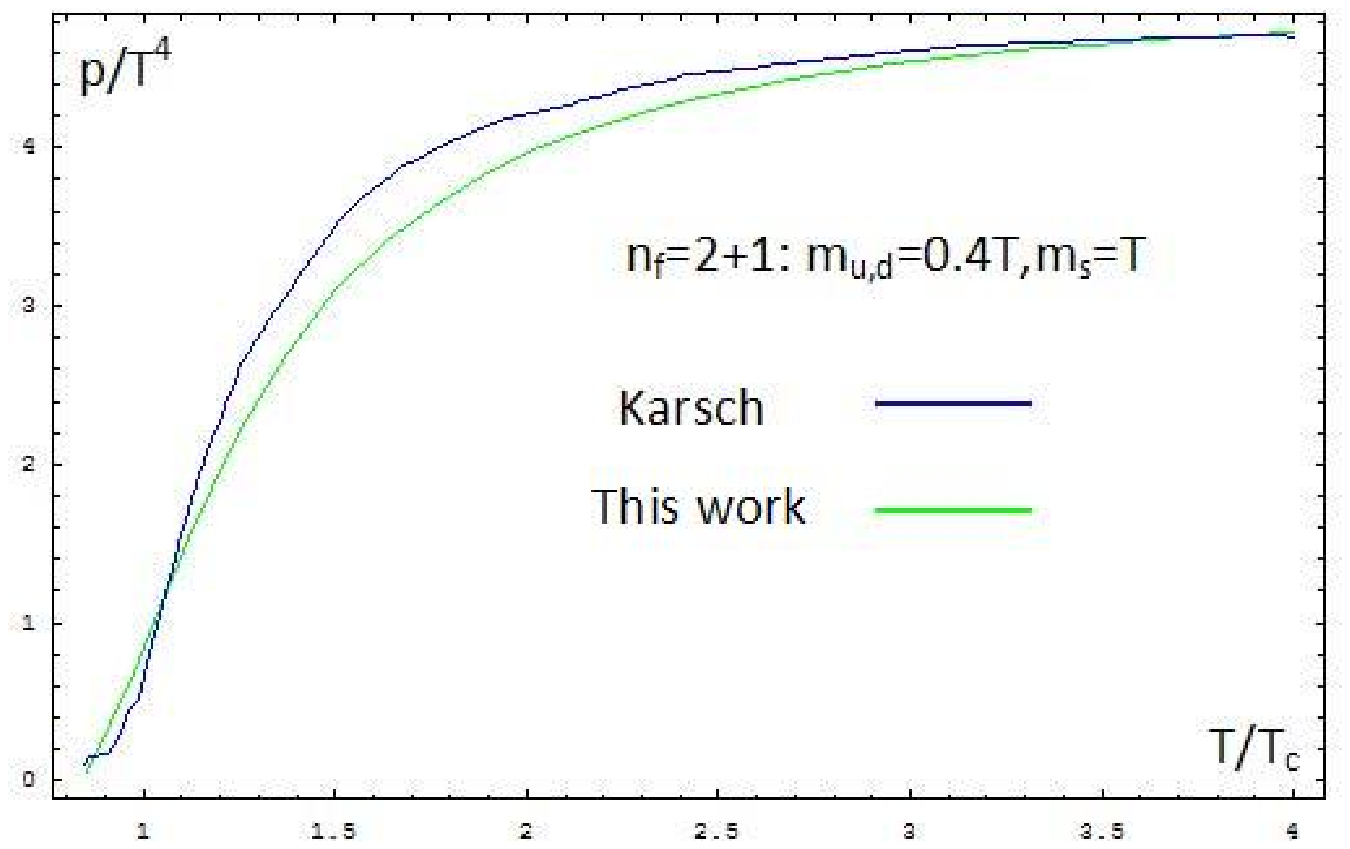}
\includegraphics[width=6.8cm]{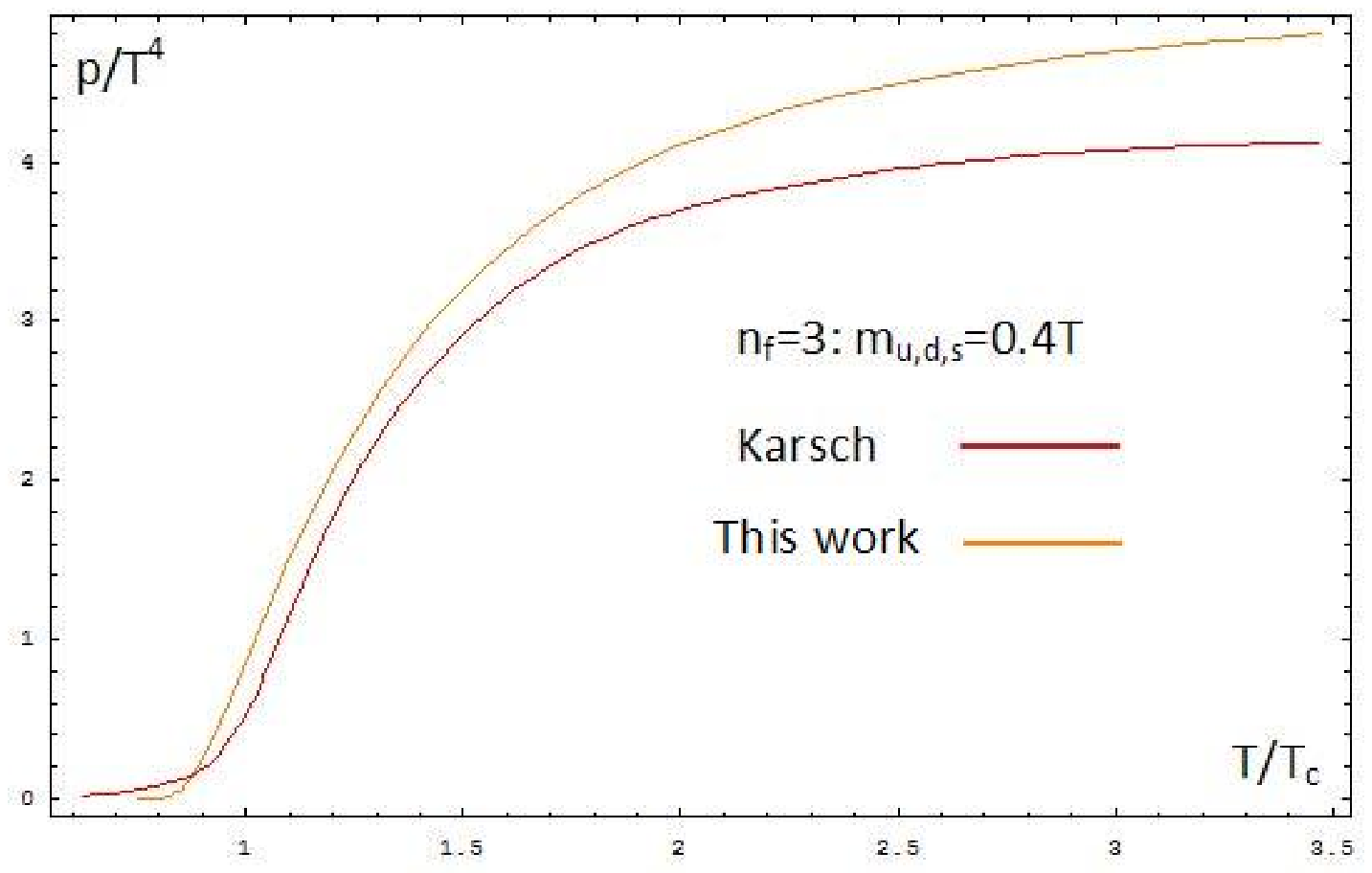}
\caption{Pressure $\frac{P}{T^4}$ as function of temperature  $T$.
The case of $n_f=2+1$ (left figure) and $n_f=3$ (right figure)
(\ref{48}),(\ref{49}). Lattice results were taken from
\cite{0701210}.}
\end{figure}

In this section we shall exploit the reduced pressure
$p=\frac{P}{T^4}$, which for $\mu > 0$ can be written as:

\begin{equation}
p_q\equiv\frac{P_q^{SLA}}{T^4}=\frac{4N_c n_f}{\pi^2}
\sum_{n=1}^{\infty} \frac{(-1)^{n+1}}{n^4} L_{fund}^n
\varphi_q^{(n)} \cosh{\frac{\mu n}{T}} \label{44}
\end{equation}

\begin{equation}
    p_{gl}=\frac{P^{SLA}_{gl}}{T^4}=\frac{2(N_c^2-1)}{\pi^2}\sum_{n=1}^{\infty}\frac{1}{n^4}L_{adj}^n\label{45}
\end{equation}

with $\varphi_q^{(n)}$ given in (\ref{25})
\begin{equation}
    \varphi_q^{(n)} = \frac{n^2 m_q^2}{2 T^2} K_2\left(\frac{nm_q}{T}\right)\label{46}
\end{equation}

Both sums can be written in a more convenient way. Using the
representation of $K_2$,
\begin{equation}
    \varphi_q^{(n)} =\frac{n^4}{6}\int_{0}^{\infty} \frac{z^4}{\sqrt{z^2+\nu^2}}e^{-n \sqrt{z^2+\nu^2}}dz\label{47}
\end{equation}
  where $\nu=m_q/T$, one has\footnote{The form (\ref{48}) was independently obtained by N.O. Agasian (to be published).}:

\begin{equation}
p_q=\frac{N_c}{3}\frac{n_f}{\pi^2}\left[\Phi_{\nu}\left(a_q-\frac{\mu}{T}
\right)+\Phi_{\nu}\left(a_q+\frac{\mu}{T}\right)\right]\label{48}
\end{equation}

\begin{equation}
p_{gl}=\frac{N_c^2-1}{3\pi^2}\int_{0}^{\infty}\frac{z^3
dz}{e^{z+a_{gl}}-1}\label{49}
\end{equation}
    with $a_q=V_1(T)/2T$, $a_{gl}=\frac{9}{4} a_q$ and

\begin{equation}
    \Phi_{\nu}(a)=\int_0^{\infty} \frac{z^4}{\sqrt{z^2+\nu^2}}\frac{dz}{e^{\sqrt{z^2+\nu^2}+a}+1}\label{50}
\end{equation}

In the paper we consider  the case of $\mu=0$ and characteristic
temperature region of $T\approx T_c$ ($T_c=170\div270$ MeV) where
quark masses do not affect the thermodynamical functions
appreciably. This is due to the fast convergence of the sum over
$n$ at large $n$ ensured by factors $\frac{1}{n^4}$, $L^n$ ($L<1$)
while $\varphi^{(n)}_q\approx 1$ for $n\approx 1$.
Characteristically, $\varphi_q^{(n)} (m_q=0) =1$, and for $m_q=0.4
T$  one has $\varphi_q^{(1)}=0.96, \varphi^{(15)}_q=0.03$.
Therefore one can with a good accuracy neglect masses in
(\ref{48}),(\ref{49}):
\begin{equation}
    p_q=\frac{2n_f}{\pi^2}\int_0^{\infty} \frac{z^3 dz}{e^{z+a_q}+1}\label{51}
\end{equation}
\begin{equation}
    p_{gl}=\frac{8}{3\pi^2}\int_0^{\infty} \frac{z^3 dz}{e^{z+a_{gl}}-1}\label{52}
\end{equation}

Eqs. (\ref{51}),(\ref{52}) are compared with lattice pressure data
in Fig.2 for $n_f=2+1$ (left) and $n_f=3$ (right figure). In Fig.3
are shown our calculated curves for the cases $n_f=2+1$ (left
part) and $n_f=3$ (right part), which are compared with lattice
data from \cite{0701210}.

To simplify further one can use for $\mu=0$ instead of
(\ref{51}),(\ref{52}) the first terms of expansion in
(\ref{44}),(\ref{45}), namely: \be p_q =\frac{12 n_f}{\pi^2}
L_{fund}\label{53}\ee \be p_g=\frac{16}{\pi^2} L_{adj}
\label{54}\ee Another useful quantities to compare with lattice
data are the internal energy density and the ``nonideality'' of
the QGP:

\begin{equation}
    \varepsilon=T^2 \frac{\partial}{\partial T}\left(\frac{P}{T}\right)_V=\varepsilon_q+\varepsilon_{gl}\label{55}
\end{equation}

\begin{figure}[!h]
\includegraphics[width=7cm]{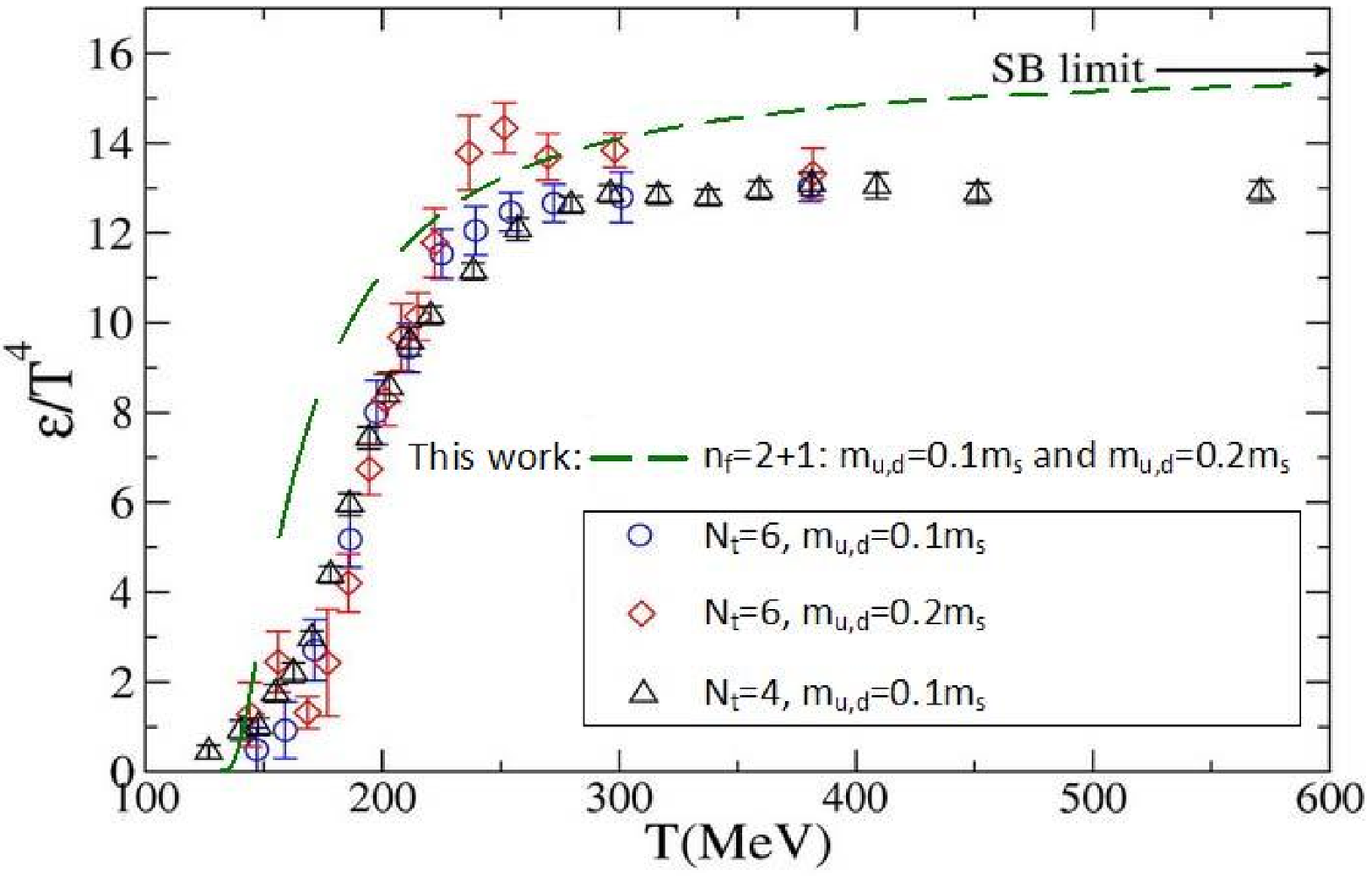}
\includegraphics[width=7cm]{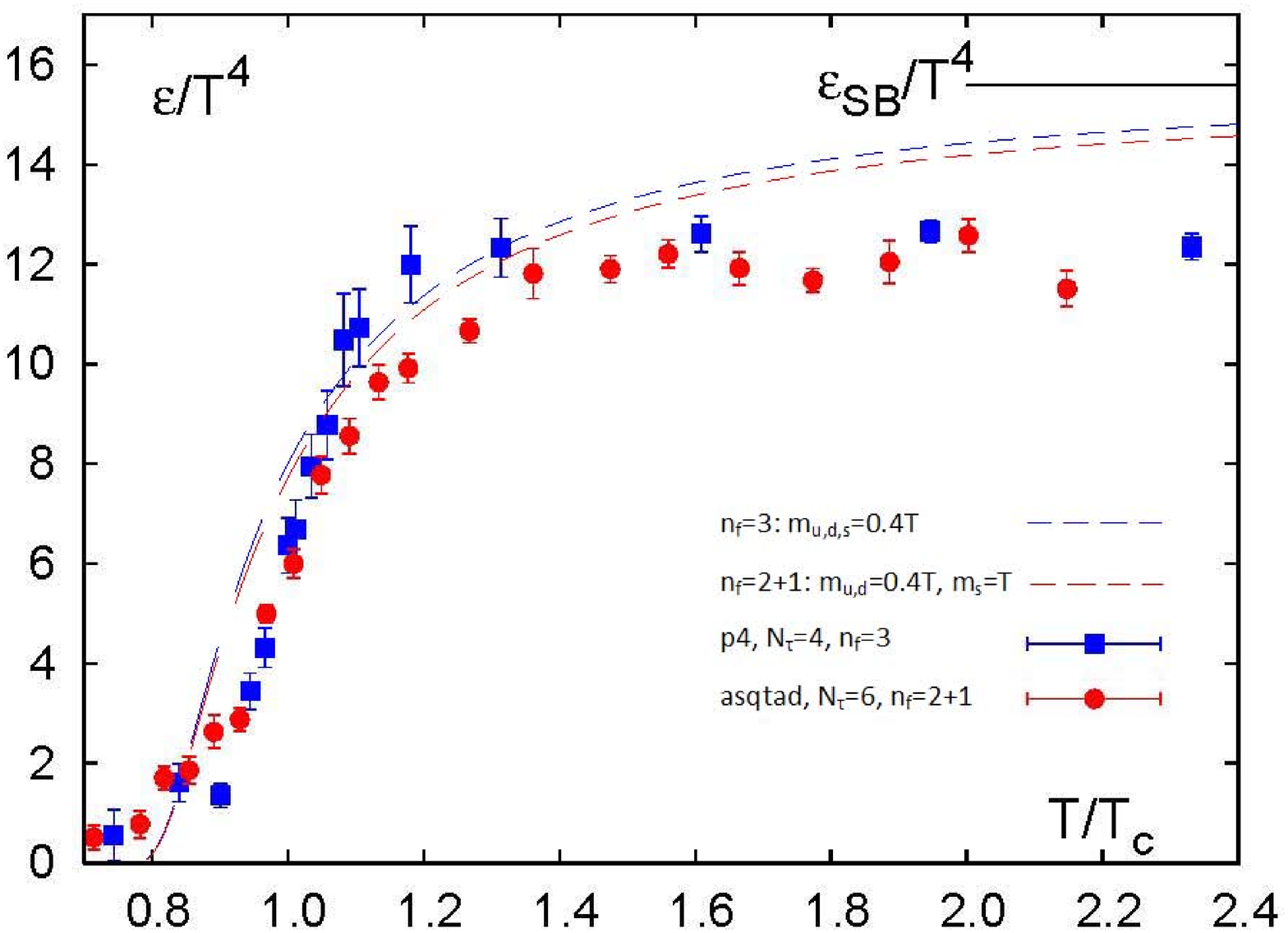}
\caption{Energy density $\frac{\varepsilon}{T^4}$ as function of
temperature  $T$. The case  $n_f=2+1$ with $m_{u,d}=0.1m_s$ and
$m_{u,d}=0.2m_s$ (green dashed curve) (\ref{55}) is compared to
lattice data from \cite{0610017}(left fig.). The case $n_f=2+1$
with $m_{u,d}=0.4T$, $m_s=T$ (red dashed curve) and $n_f=3$ with
$m_q=0.4T$ (blue dashed curve) (\ref{55}) are compared to lattice
data from \cite{0608003}(right fig.).}
\end{figure}

\begin{figure}[h]
\includegraphics[width=12cm]{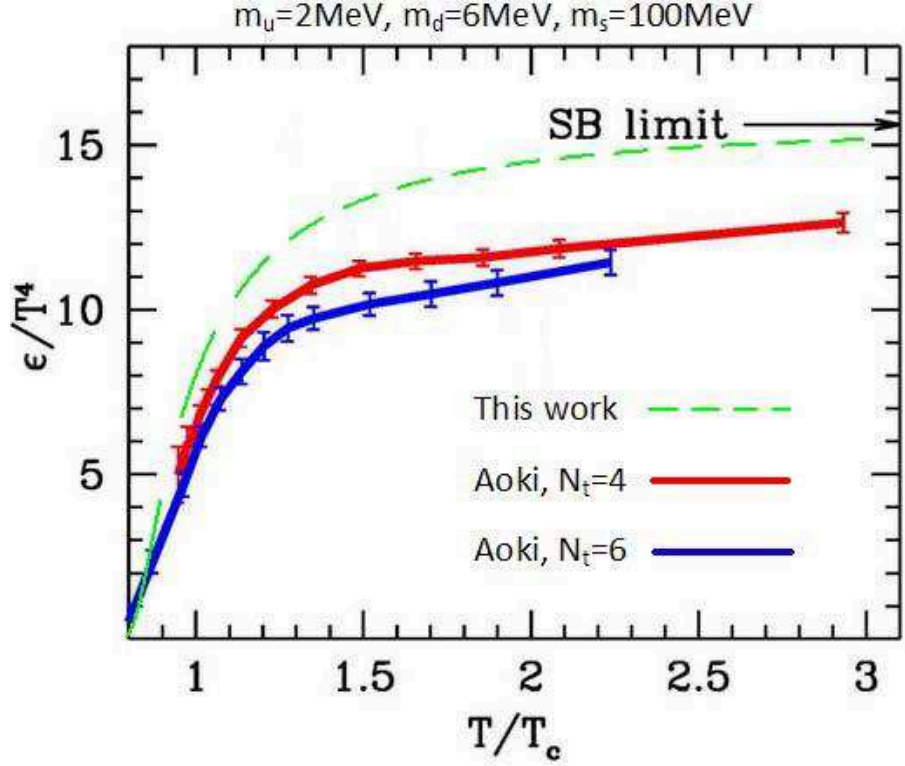}
\caption{Energy density $\frac{\varepsilon}{T^4}$ as function of
temperature  $T$. The curve for $n_f=3$ with
$m_{u}=2$MeV,$m_{d}=6$MeV,$m_{s}=100$MeV (green dashed) (\ref{55})
is compared to lattice data from \cite{0510084}.}
\end{figure}

Using (\ref{48}),(\ref{49}) one has
\begin{equation}
    \varepsilon^{(0)}_{q}=\sum_{n_f}\frac{2}{\pi^2}T^2\frac{d}{dT}\left(T^3 \int_0^{\infty} \frac{z^4}{\sqrt{z^2+\nu^2}}\frac{dz}{e^{\sqrt{z^2+\nu^2}+a_q}+1}\right)\label{56}
\end{equation}

\begin{equation}
    \varepsilon^{(0)}_{gl}=\frac{3}{3\pi^2}T^2\frac{d}{dT}\left(T^3 \int_0^{\infty} \frac{z^3dz}{e^{z+a_{gl}}+1}\right)\label{57}
\end{equation}

 and the ``nonideality'' of the QGP:
\begin{equation}
    I(T)=\frac{\varepsilon-3P}{T^4}=T \frac{\partial p}{\partial T}\label{58}
\end{equation}

In the simple approximation (\ref{53}),(\ref{54}) one has \be
I(T)=\frac{12
n_f}{\pi^2}T\frac{dL_{fund}}{dT}+\frac{16}{\pi^2}T\frac{dL_{adj}}{dT}\label{59}\ee

We compare our calculations for $\frac{\varepsilon}{T^4}$ in Fig.4
and 5 with three different lattice data:
\cite{0608003},\cite{0610017},\cite{0510084}. In Fig.6 we
demonstrate our $I(T)$ computed from
(\ref{58}),(\ref{51}),(\ref{52}) with lattice data of 2+1 flavor
from \cite{0510084} (left curve) and from \cite{0610017} (right
curve).

\begin{figure}[!h]
\includegraphics[width=6.7cm]{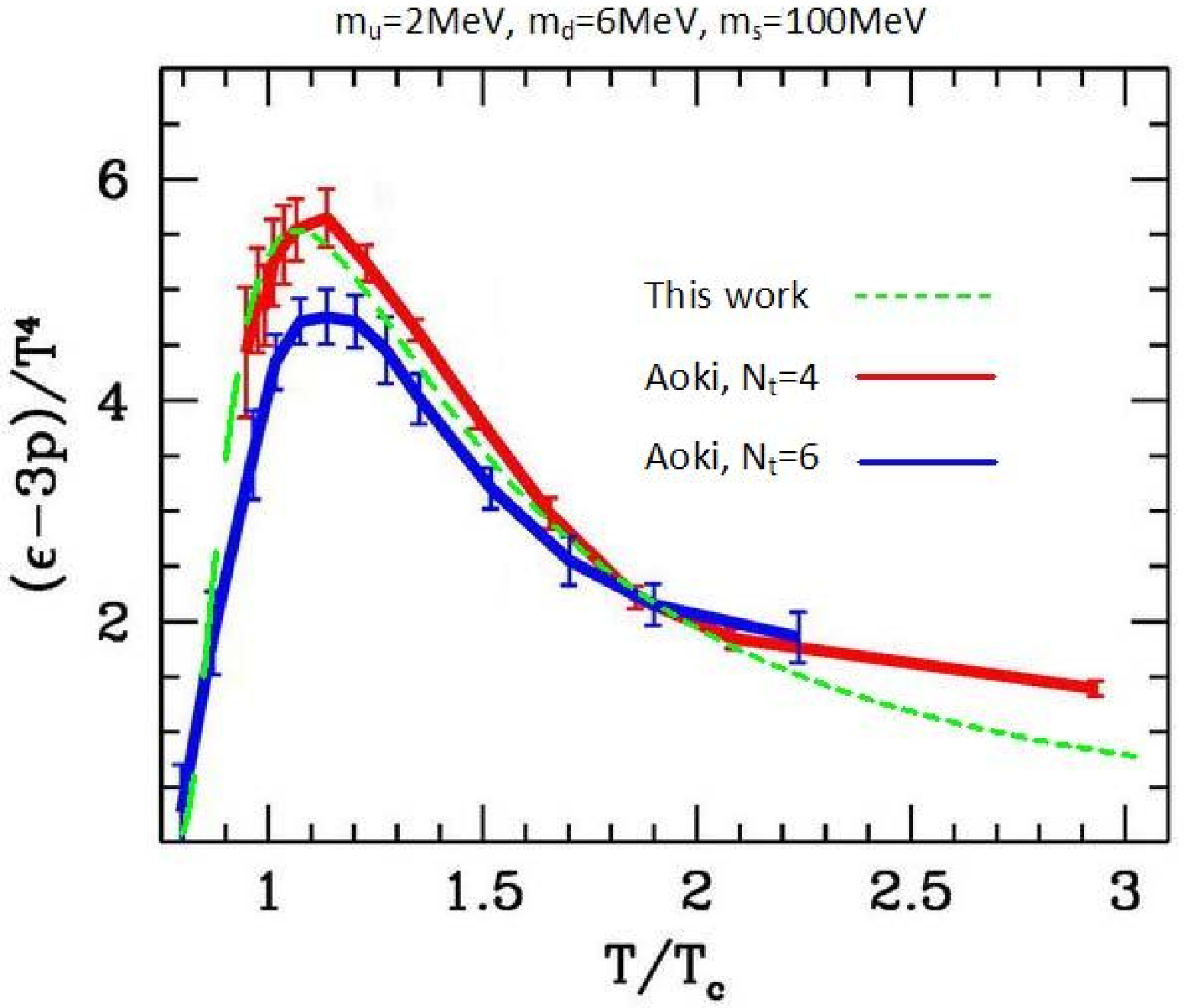}
\includegraphics[width=7.3cm]{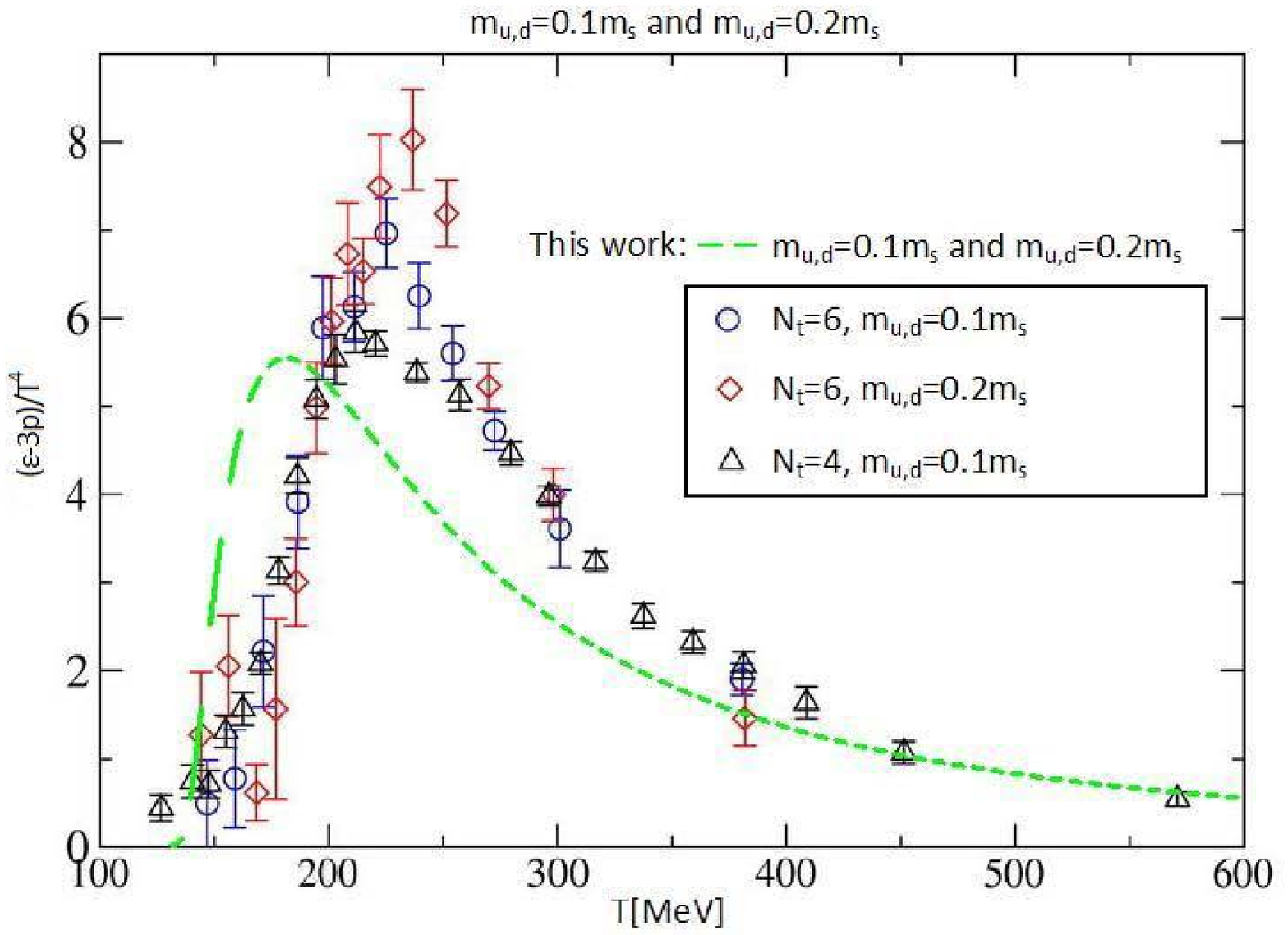}
\caption{"Nonideality" of QGP $(\varepsilon-3p)/T^4$. Shown are
the curves for (left fig.) $n_f=3$ with $m_u=2$ MeV, $m_d=6$ MeV,
$m_s=100$ MeV (green dashed line) compared to \cite{0510084} and
(right fig.) for $n_f=2+1$ with $m_{u,d}=0.1m_s$ and
$m_{u,d}=0.2m_s$ compared to \cite{0610017}. Analytical
calculations are done using (\ref{48}),(\ref{49}),(\ref{55}).}
\end{figure}

At this point it is instructive to estimate the contribution of $q
\bar q$, $gg$ interactions to the pressure. Writing the virial
coefficient in the form
$P_j=P_j^{(0)}(1+\frac{P_j^{(0)}}{T}B_j(T)+\ldots)$, where
$P_j^{(0)}=P_q,P_{gl}$ in SLA, Eqs. (\ref{21}),(\ref{22}), with
\be B_j(T)=\frac{1}{2}\int (1-e^{U_j(r,T)/T})dV, j=\mbox{fund,adj}
\label{60}\ee and taking for $q\bar q$ and $gg$ interaction term
$U_{fund}$ and $U_{adj}$ respectively at large $T$ as $U_j(r,T)=T
u_j(rT)$, one obtains a corrected pressure \be
P=P_q^{(0)}(1-c_q)+P_{gl}^{(0)}(1-c_{gl})\label{61}\ee where
$c_{gl}\cong \frac{16}{\pi^2}\int_0^\infty \rho^2 d\rho
(e^{|u_{adj}(\rho)|}-1)$, $c_{q}\cong \frac{12
n_f}{\pi^2}\int_0^\infty \rho^2 d\rho (e^{|u_{fund}(\rho)|}-1)$.
 Note that $q\bar q$ and $gg$ interaction in the singlet color
 state is attractive, so that $|U_j|=-U_j$. The dependence on $rT$
 in $u_j$ occurs at large $T$, in the dimensionally reduced
 regime, when dynamical dimensional quantity is the spatial string
 tension $\sigma_H=\mbox{const}\cdot T^2$, and the Debye mass $m_D(T)\cong
 2\sqrt{\sigma_H}=\mbox{const}\cdot T$.

 Thus one expects that 1) the corrected pressure is smaller than
 the SLA predicts, 2) the large $T$ behavior of $P(T)$ is below
 the Stefan-Boltzmann values (modulo logarithmic factors). Both
 features are clearly seen in the Fig.2,3,4,5.

\section{Discussion of results. Conclusions.}

We have shown in section 2, following \cite{2}, that EoS in the
zeroth approximation is represented by free quark and gluon lines
augmented by the factors $L_{fund}$ for quarks and $L_{adj}$ for
gluons. These factors have been derived from the Gausssian
color-electric correlators $D^E(x)$, $D^E_1(x)$, and the latter in
its turn can be computed analytically from the gluelump Green's
function, or directly on the lattice \cite{8,9}. This
representation of $L_{adj}$ and $L_{fund}$ allows to express
$L_j$, $j=fund, adj$ in terms of the NP static potential
$V_1(r,T)$ at $r=\infty$, and compare the latter with the singlet
free energy $F_{Q\bar Q}^1(r,T)$. It was argued that $V_1$ and
$F_{Q\bar Q}^1$ differ due to presence of excited $Qg^n$ states in
$F_{Q\bar Q}^1$, and can be taken equal in the first
approximation. This leads to the identification of $L_j$ with the
modulus of corresponding Polyakov lines. In the Gaussian
approximation for $V_1$ one then automatically obtains the Casimir
scaling for $L_j$: $L_{adj}=(L_{fund})^\frac{C_2(adj)}{C_2(fund)}$
which is observed on the lattice with good accuracy \cite{4}.
Corrections are found to be of two types: 1) contribution of
higher correlators to $V_1$ and $L_j$ yields less then 10\%
(\ref{20}) and can be neglected 2) contribution of excited states
yields corrections not connected by Casimir scaling and therefore
high accuracy of data \cite{4} imposes a stringent limit on the
role of excited states of the type $(Qg^n)$. For $T>T_c$ our
expression (\ref{29}) automatically predict vanishing of
$L_{fund}$ for $n_f=0$ and behavior of $L_{adj}\cong
\exp{(-M_0/T)}$ with $M_0$ - lowest gluelump mass $\approx 1$ GeV.
These features are in good agreement with the lattice data
\cite{4}, and are shown in Fig.1.

For EoS using formulas of section 2 and treatment in \cite{2} we
have given two types of expressions for the pressure $P$: 1) as a
sum over winding $n$ (Matsubara frequencies) in (\ref{44}),
(\ref{45}) and equivalent forms as integrals over "momentum" $z$
in (\ref{48}),(\ref{49}). It was argued that for $\mu=0$ and not
large $T$, $T_c\leqslant T \lesssim 2T_c$  one can use much
simpler forms of (\ref{53}),(\ref{54}), which are first terms of
the sums (\ref{44}),(\ref{45}).

In all these forms the only source of non-perturbative dynamics in
EoS is Polyakov factors $L_j$, which are defined independently and
therefore our EoS is the explicit prediction without any model of
fitting parameters. Hence check of our approach is the check of
our basic principle that non-perturbative dynamics enters in the
form of vacuum based factors $L_j$.

Comparison of our EoS, (\ref{44}),(\ref{45}) or
(\ref{48}),(\ref{49}), is done with several lattice groups for
each quantity, to have an idea of accuracy of our results and of
lattice data, and dependence on quark masses. The latter appears
very weak in EoS, e.g. quark mass of $m_q=0.4T$ yields a 4\%
correction to the zero mass result, while on the lattice this
dependence is stronger. We compare pressure $\frac{P}{T^4}$ for
$n_f=0,2,3$ and $m_q=0.4T$ in Fig.3 (left part). One can see
deviation of $\sim 20\%$ of our curves from lattice data
\cite{0608003} for $T\lesssim 3T_c$ and the same type of agreement
for $n_f=2+1$ with data from \cite{0610017}. Typically our curves
are higher with the fact that the (attractive) interaction between
quarks, antiquarks and gluons is not taken into account. The first
correction (\ref{60}),(\ref{61}) treating this attraction between
$q\bar q$ and $gg$, has the negative sign, which might improve the
agreement. The agreement is however better with another set of
lattice data from \cite{0701210} done for $n_f=2+1$, see Fig.3.
Comparing left and right parts of the Fig.3 one can notice, that
lattice data \cite{0701210} are much more sensitive to the quark
masses, than our prediction.

Another interesting comparison is for the internal energy
$\varepsilon$ and non-ideality $I=\frac{\varepsilon-3P}{T^4}$,
given in Fig.4,5 and 6. It is important that both quantities
contain derivatives $T\frac{dL_j(T)}{dT}$ and therefore are much
more sensitive to the type of non-perturbative dynamics, which is
present in our approach. The agreement of our Eqs.
(\ref{56}),(\ref{57}) with data from \cite{0608003} and
\cite{0610017} are shown in Fig.4 and is of the same quality as
for the pressure: one has $\sim 15\%$ higher theoretical curve for
$T>1.2 T_c$, the same one can see in Fig.5 with data from
\cite{0510084}. Note, that the quark masses in this case are close
to physical ones. Finally, the non-ideality is compared to the
data from \cite{0510084} in the left part of Fig.6 and is in good
agreement with data \cite{0610017} (right part of Fig.6 is less
successful, because lattice data from \cite{0610017} and
\cite{0510084} differ strongly). As a whole, it is surprising that
such simple approach without any parameters (actually primitive
formulas (\ref{53}),(\ref{54}) already have sufficient accuracy
within our approximation) yields a reasonable agreement with
lattice data for $P(T)$, $\varepsilon(T)$ and $I(T)$. If one adds
to that a good agreement of our phase curve $T_c(\mu)$ in \cite{5}
with majority of lattice data, the possible conclusion might be,
that our zeroth approximation to the non-perturbative vacuum
fields - taking non-perturbative contribution in the form of $L_j$
- is a viable spproach to the dynamics of QGP. The next step is an
account of possible perturbative and non-perturbative interactions
between quarks, antiquarks and gluons, which is partly done in
\cite{3,9} for color-electric fields $(V_1(r,T))$ and in \cite{20}
for color-magnetic ones. The exact contribution of these effects
to the EoS is not yet done and should be an important next step.
The strong interaction in $(q\bar q)$ and $(gg)$ systems
discovered in \cite{20} might give further support for the idea of
strong quark-gluon plasma - sQGP.

The authors are indebted for useful discussions to members of ITEP
and FIAN physical seminars. We are also grateful to S.N.Fedorov
for providing us with a useful program "GetData". The financial
support of the RFFI grant 06-02-17012 and NSh-843.2006.2 is
acknowledged. This work was supported by the Federal Agency for
Atomic Energy of Russian Federation.



\end{document}